\documentclass[RNAAS]{aastex62}

\usepackage{CJK}
\usepackage{graphicx}

\received{\today}
\revised{\today}
\accepted{\today}
\submitjournal{RNAAS}

\shorttitle{Supernova Sussed Serendipitously}
\shortauthors{Guillochon et al.}

\begin{document}
\begin{CJK*}{UTF8}{gbsn}

\title{Serendipitous Discovery of a 14-year-old Supernova at 16 Mpc}

\correspondingauthor{James Guillochon}
\email{guillochon@gmail.com}

\author[0000-0002-9809-8215]{James Guillochon}
\affiliation{Harvard Center for Astrophysics, 60 Garden St., Cambridge, MA 02138, USA}

\author[0000-0001-5266-670X]{Jorge Stockler de Moraes}
\noaffiliation

\author[0000-0002-2555-3192]{Matt Nicholl}
\affiliation{Harvard Center for Astrophysics, 60 Garden St., Cambridge, MA 02138, USA}

\author[0000-0002-7507-8115]{Daniel J. Patnaude}
\affiliation{Harvard Center for Astrophysics, 60 Garden St., Cambridge, MA 02138, USA}

\author[0000-0002-4449-9152]{Katie Auchettl}
\affiliation{Center for Cosmology and AstroParticle Physics (CCAPP), The Ohio State University, 191 W. Woodruff Ave., Columbus, OH 43210, USA}
\affiliation{Department of Physics, The Ohio State University, 191 W. Woodruff Avenue, Columbus, OH 43210, USA}
\nocollaboration

\author[0000-0002-3026-0562]{Aaron J. Barth}
\affiliation{Department of Physics and Astronomy, University of California, Irvine, 4129 Frederick Reines Hall, Irvine, CA 92697, USA}

\author[0000-0001-6947-5846]{Luis C. Ho}
\affiliation{Kavli Institute for Astronomy and Astrophysics, Peking University, Beijing 100871, P. R. China}
\affiliation{Department of Astronomy, School of Physics, Peking University, Beijing 100871, China}

\author[0000-0001-5017-7021]{Zhao-Yu Li (李兆聿)}
\affiliation{Key Laboratory for Research in Galaxies and Cosmology, Shanghai Astronomical Observatory, Chinese Academy of Science, 80 Nandan Road, Shanghai 200030, China}
\affiliation{College of Astronomy and Space Sciences, University of Chinese Academy of Sciences, 19A Yuquan Road, Beijing 100049, China}
\affiliation{LAMOST Fellow}
\collaboration{Carnegie-Irvine Galaxy Survey}


\section{Introduction}

In this Research Note we present a serendipitous discovery of the transient CGS2004A (AT2004iu)\footnote{\url{https://sne.space/sne/CGS2004A}} in an image collected by the Carnegie-Irvine Galaxy Survey of the Scd: \citep{deVaucouleurs:1991a} galaxy NGC~1892, which we determine is most likely to have been a type IIP supernova.

\section{Data}

\begin{figure*}
\centering
\begin{picture}(488,390)
\put(0,180){\includegraphics[clip,width=0.47\linewidth]{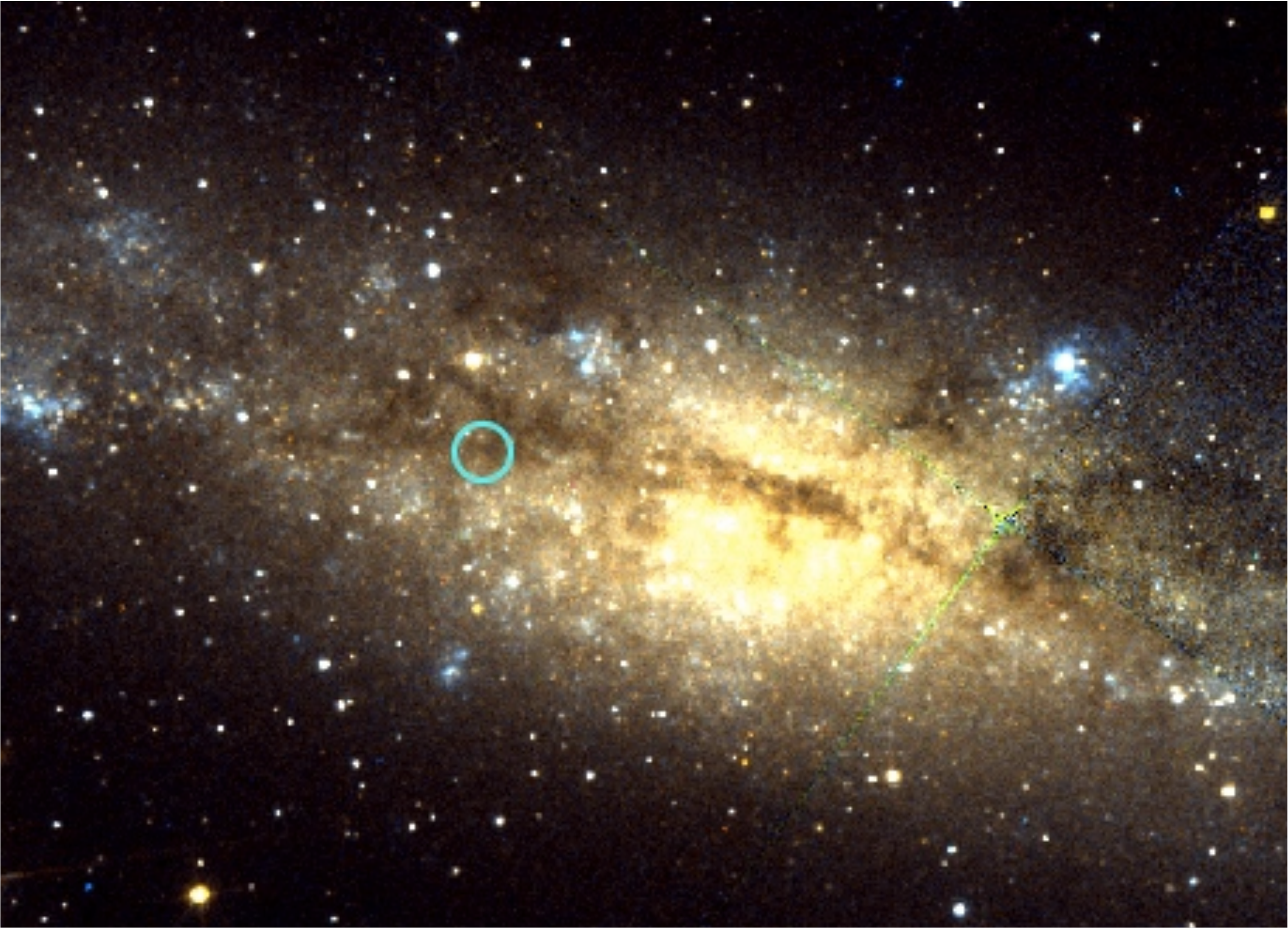}}
\put(80,185){{\color{white} \bf{HST -- 2001-09-22}}}
\put(247,180){\includegraphics[clip,width=0.47\linewidth]{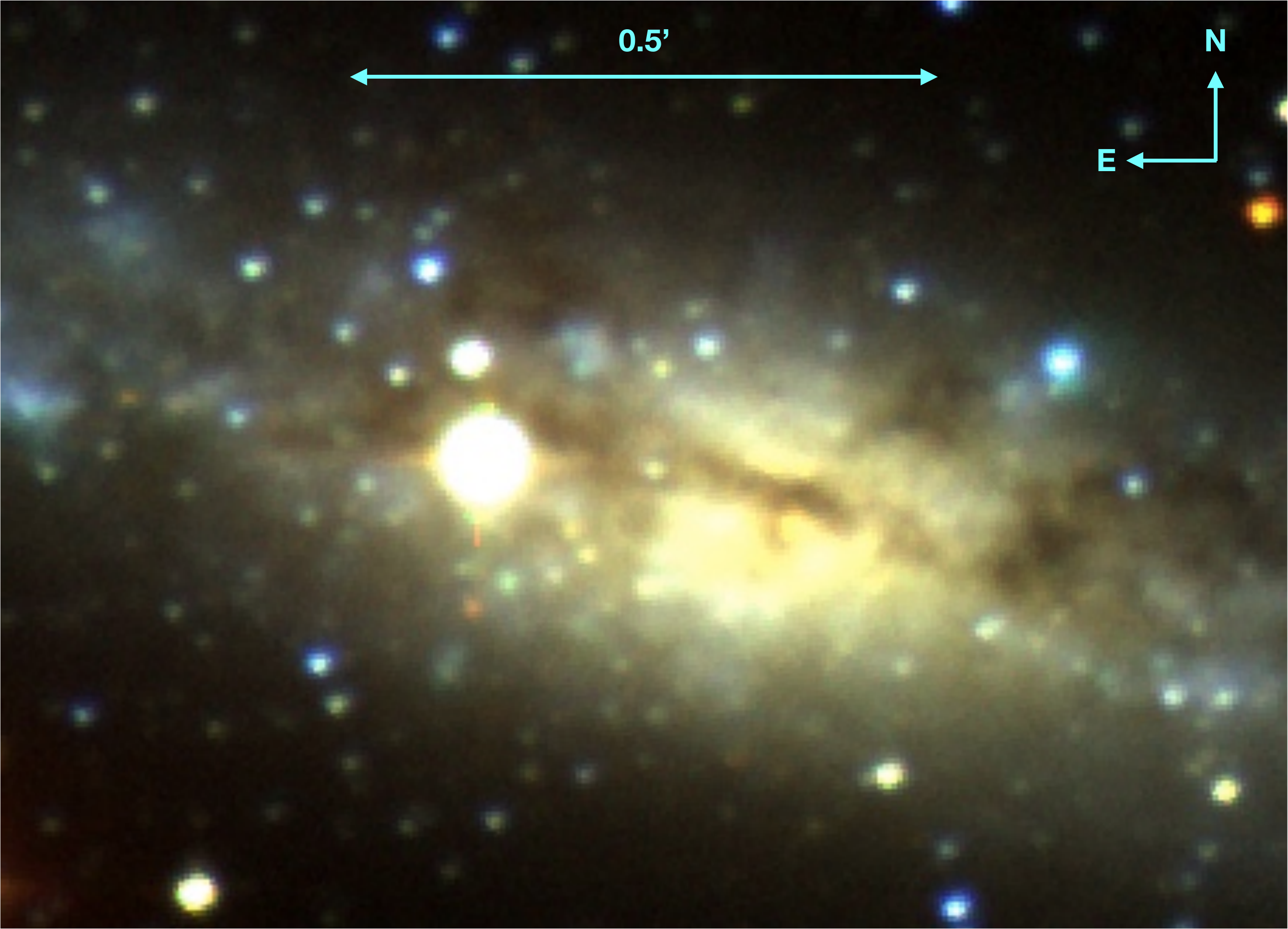}}
\put(324,185){{\color{white} \bf{CGS -- 2004-01-20}}}
\put(-3,0){
\includegraphics[width=0.47\linewidth]{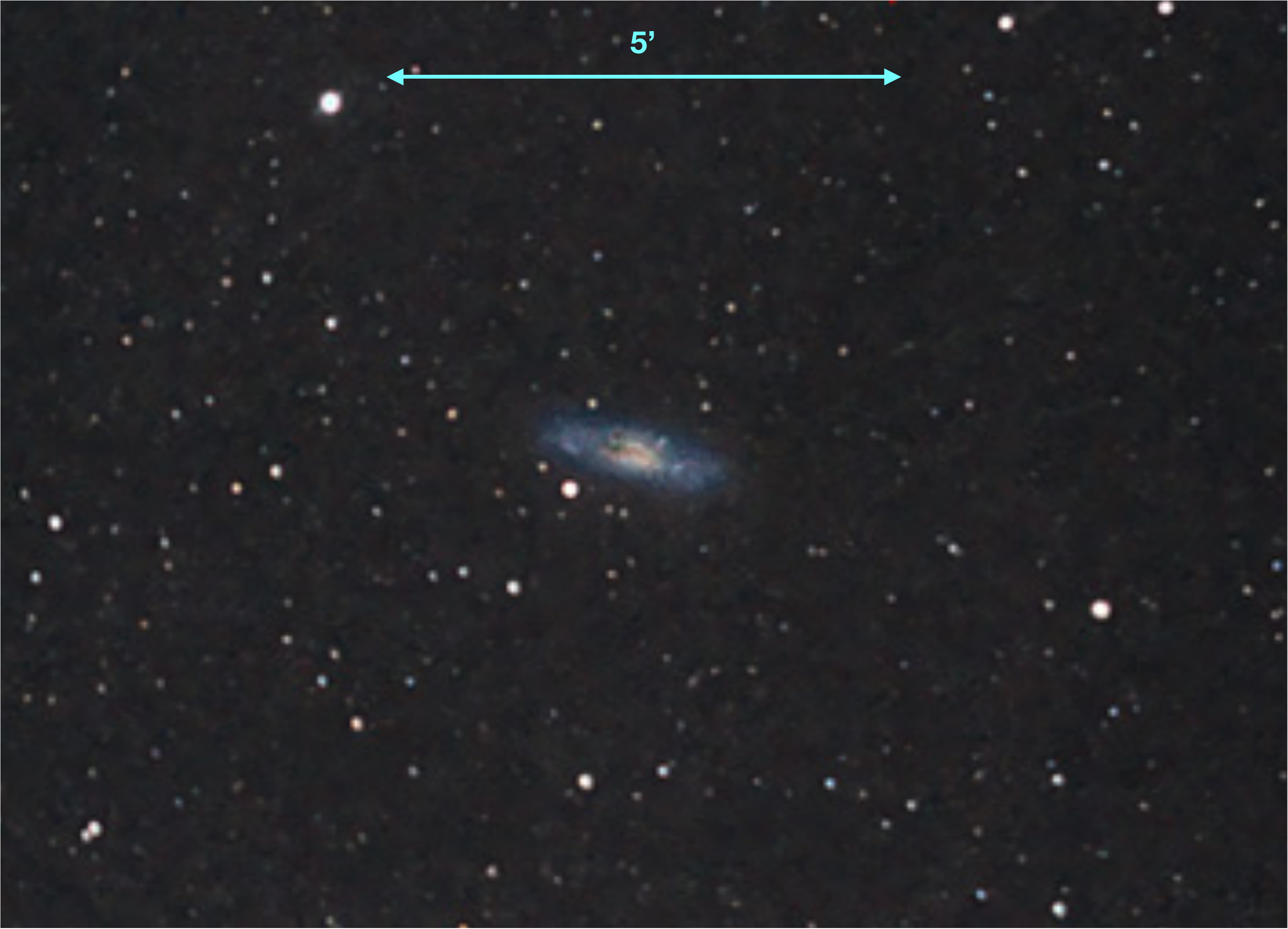}}
\put(40,5){{\color{white} \bf{J. Stockler de Moraes -- 2017-01-28}}}
\put(247,0){\includegraphics[width=0.47\linewidth]{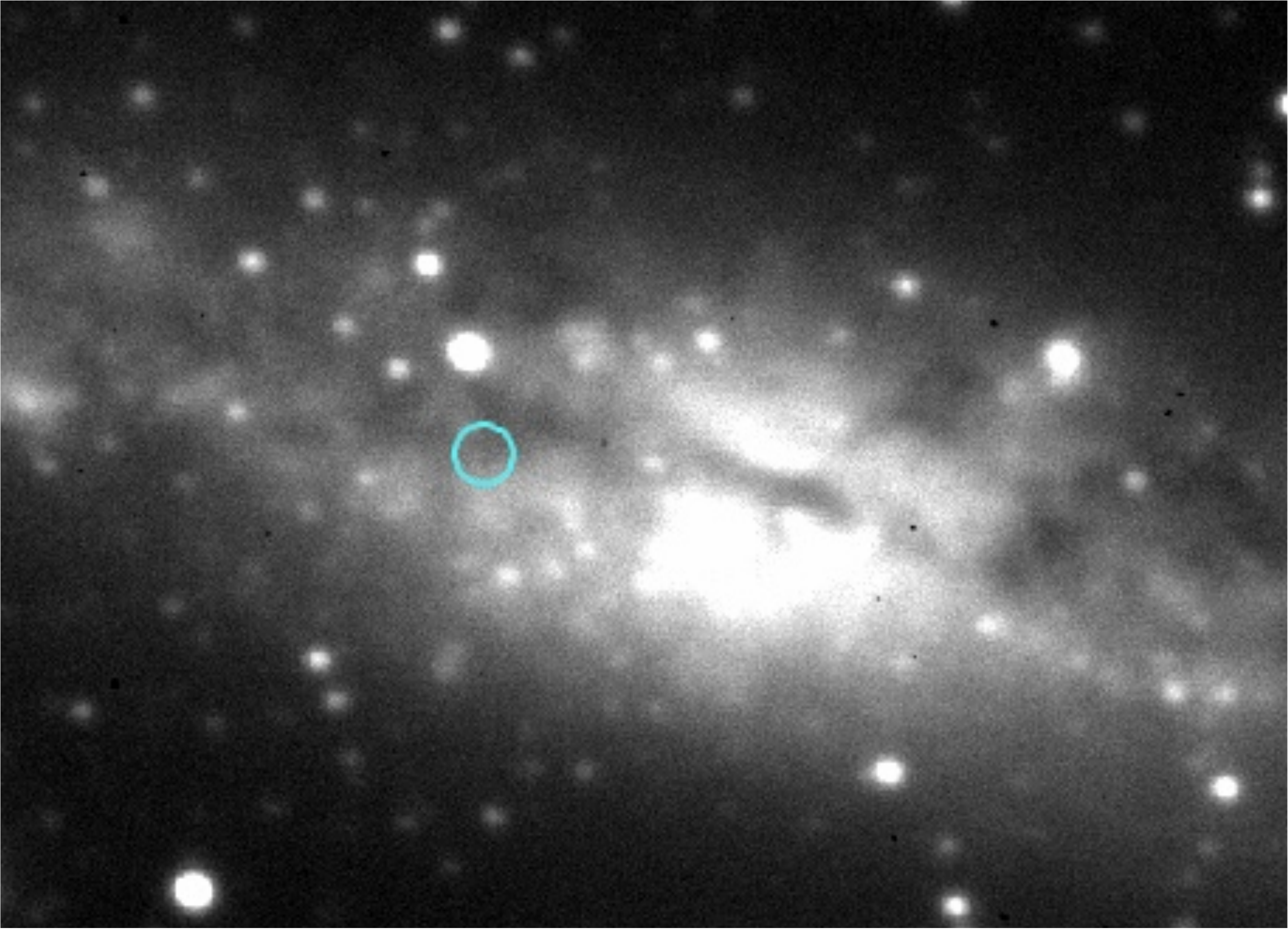}}
\put(314,5){{\color{white} \bf{Magellan -- 2018-08-06}}}
\end{picture}
\caption{Images collected before (top left corner, \textit{HST}, combined F450W/ F814W image), during (top right corner, CGS survey, combined \emph{BVI} image), and after (bottom left and right images, Stockler de Moraes and Magellan IMACS \emph{r}-band, respectively) the transient (compiled by author Nicholl). The \textit{HST}, CGS, and Magellan images are aligned into the same coordinate system/scale, whereas the bottom left image shows a wider-frame image captured by author Stockler de Moraes (scale shown by cyan bar). The cyan reticle in the \textit{HST} and Magellan images (diameter: 3 arcseconds) shows the position of the source.}\label{fig:images}
\end{figure*}
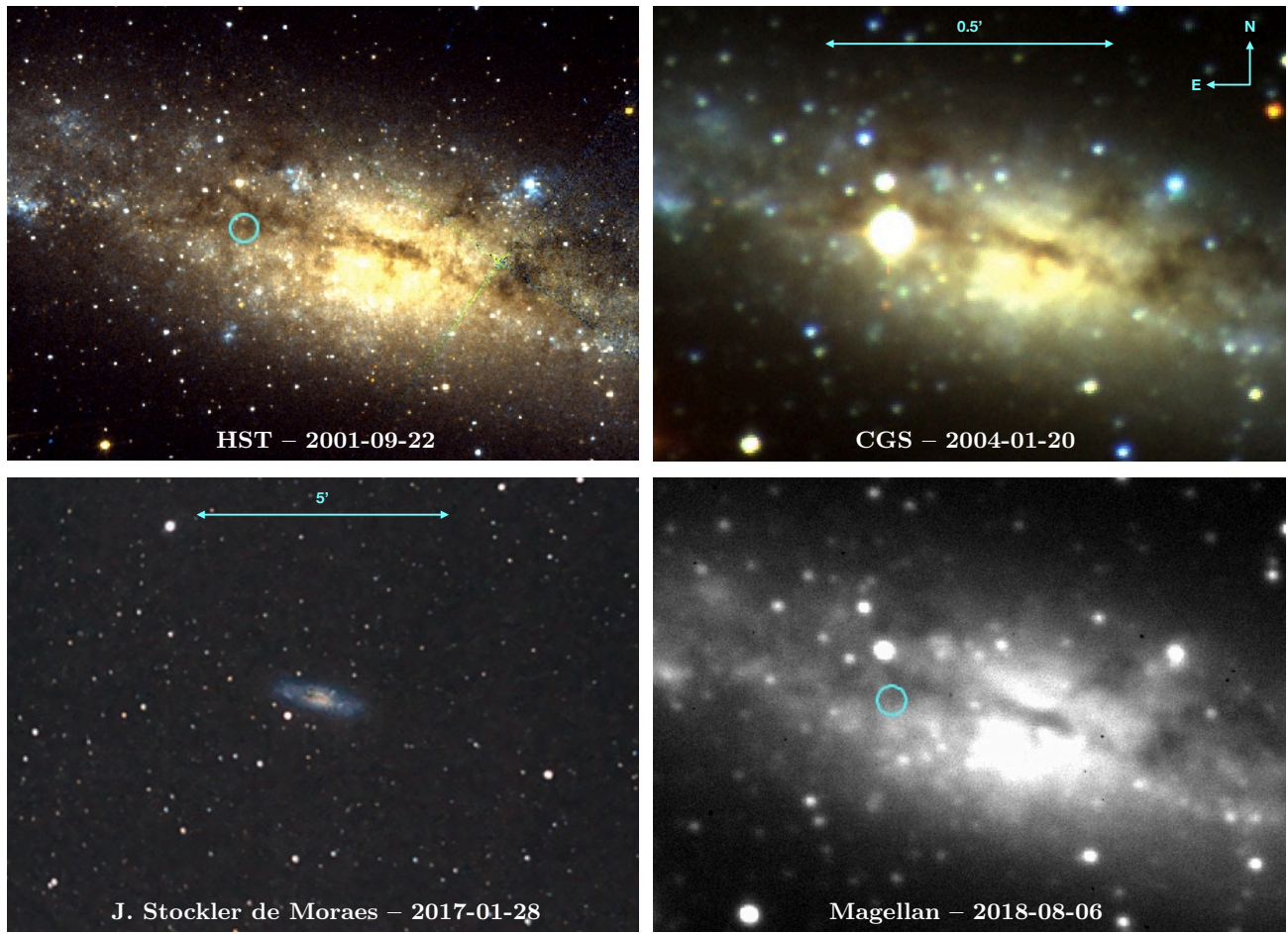

Images of NGC~1892 are presented in Figure~\ref{fig:images}. Imaging was performed on 28 January, 2017 by author Stockler de Moraes using a Canon 1100D mounted to a GSO 305 mm telescope with a GSO coma corrector and Skyglow filter from the Observatory of Vilatur, with a pixel scale of 0\farcs6. Author Stockler de Moraes compared his image to a published image of the galaxy presented by the CGS\footnote{\url{https://cgs.obs.carnegiescience.edu/CGS/object_html_pages/NGC1892.html}}, collected on 20 January, 2004 \citep{Ho:2011a}. A bright source whose final position was determined to be 05:17:11.53, $-$64:57:31.6 was noted in the Carnegie-Irvine Galaxy Survey (CGS) image, at which point author Guillochon was contacted. Author Guillochon determined that no known point source is cataloged within an arcminute of the source\footnote{\url{https://simbad.u-strasbg.fr/}}\textsuperscript{,}\footnote{\url{https://sne.space}}, nor is any minor planet identified at the location at the time of the CGS observation\footnote{\url{https://www.minorplanetcenter.net/}}.

Aperture photometry was performed on the CGS image by authors Nicholl and Barth, establishing that the source was saturated in all four bands (\emph{BVRI}). However, the \emph{B}-band image was found to be only modestly saturated, allowing an apparent magnitude estimate of 14.6 in \emph{B} to be obtained, accounting for 0.2 magnitudes of extinction in \emph{B} from the Milky Way \citep[14.8 without extinction,][]{Schlafly:2011a}. Given the estimated host galaxy distance of 15.5~Mpc \citep{Helou:1991a}, this implies an absolute \emph{B} magnitude of $-$16.4.

We downloaded archival images of NGC~1982 from the Hubble Legacy Archive. These images date from 2001, before the CGS images, but no source is seen at that position in the \textit{Hubble Space Telescope} images.
Given the absolute magnitude of the source, and its absence in images taken before and after the CGS detection, we conclude that it was a almost certainly a supernova.

Follow-up \emph{r}-band imaging was conducted by author Patnaude with IMACS on the Baade telescope at Magellan to look for potential late-time emission. No detectable emission was found at the source location.

\section{Discussion}

As a galaxy with stellar mass $M_\ast = 4 \times 10^9 M_\odot$ \citep{Foord:2017a}, the fraction of type II supernovae in NGC~1892 is slightly more than half of all supernovae expected to occur within the galaxy \citep{Graur:2017a,Graur:2017b}, and given the host galaxy's star-forming nature \citep[$\simeq 1~M_\odot~{\rm yr}^{-1}$,][]{Helmboldt:2004a}, the probability that this was a core collapse event is most likely higher than this fraction. Its absolute magnitude in the CGS image, $-$16.4 in the \emph{B} band, is typical of type IIP supernovae in their plateau phase \citep{Anderson:2014a}, which lasts roughly 100 days and is thus the time at which a IIP supernova is most likely to be observed (type Ib/c and IIb supernovae are much shorter-lived).

To check the possibility that this supernova might belong to a particular II subtype that produces X-ray emission, a search was performed in the Chandra X-ray archive by author Auchettl. An observation of this galaxy was found to have been performed with the ACIS-S instrument on 23 August, 2015, from which a three-sigma upper limit of $5 \times 10^{-4}$~counts~s$^{-1}$ was estimated between 0.3 and 10 keV ($\lesssim 10^{36} ~{\rm ergs~s}^{-1}$), significantly below the count rate expected from X-ray emitting IIn, II-L, Ib/c, and IIb SNe at an age of 11 years, but comparable to some IIP SNe \citep{Dwarkadas:2014a,Ross:2017b}. Given the preponderance of evidence, the most likely scenario is that this transient was a supernova of type IIP.

\acknowledgments

We thank Iair~Arcavi, Or~Graur, Vikram~Ravi, Stuart~Ryder, Brad~Tucker, and Lukasz~Wyrzykowski for helpful input. HST imaging was obtained from the HST legacy archive, proposal ID 9042 with PI Smartt. This paper includes data gathered with the 6.5 meter Magellan and 2.5 meter du Point telescopes located at Las Campanas Observatory, Chile.

\vspace{5mm}
\facilities{HST(WFPC2), Magellan(IMACS), du Pont(2.5m)}

\software{astropy \citep{Astropy-Collaboration:2013a}, ds9 \citep{2003ASPC..295..489J}}

\bibliography{references}

\end{CJK*}
\end{document}